# Characterising Communities of Twitter Users Who Posted Vaccines Related Tweets by Information Sources

Md. Rafiul Biswas

College of Science and Engineering, Hamad Bin Khalifa University, Doha, Qatar

## Abstract

*Background: Measuring what and how information generates and propagates in online communities such as Twitter user communities make a distinguishable characteristic over the of demographical region and it helps the public health organization to take appropriate decision to develop the public health system.*

*Objective: We formed community structure based on web page credibility and we measured the types of information for characterizing communities of tweeter users who posted about tweets related to vaccine.*

*Methods: We performed the experiment on only Twitter data (tweets) regarding vaccine. The duration of data collection was between 17 January 2017 and 14 March 2018. We formulated cluster based on the information on its contents and sources it resides (i.e., website domains). We only focused the topics which were related to vaccine. To detect the structure and network of community, we applied Louvain community algorithm along with Random walks called Info map method over vaccines related tweeter user. We defined the communities based on various measures derived from the information shared by Twitter users. Representations and visualizations of the communities based on these derived measures help the public health organization to make understand the possible cause of the rejecting the safety and efficacy of vaccines.*

*Results:* To analyse people perception over social media, we downloaded 6,591,566 tweets. We distinguished 1,860,662 users who were posting related to vaccine. We applied Louvain community detection algorithm along with DMM values and we found 192 communities. It also produced higher alignment values as long as the number of topics were low. With the increase of topics number, the alignment becomes lower. We found 30 topics after applying the two algorithms and we organized the topics in a hierarchy order. The total tweets were divided into two parts based on their characterizes. The first category contained 163,148 (57.16%) tweets which were based as evidence and advocacy and the second category contained 6244 (2.19%) tweets which were based on the experiences and opinion. We observed 4548 users were posting about experiential tweets about vaccine and of them 3449 users (75.84%) were posted evidence and advocacy related.

*Conclusion:* The novel approach to detect community based on information sources (web page) credibility shared through tweets is appeared to be a convenient solution. It will help the public health organization to identify the root cause of low health outcome. Our study will also help to build a public surveillance system where it can easily identify the percentages of negative opinion as well as the percentages of positive opinion regarding vaccine.

## Keywords

# 1 Introduction

Many people are suffering and dying every year due to some diseases like COVID-19 pandemic, Hepatitis B, Influenza. The COVID-19 pandemic has changed the motion of life worldwide with the increasing number of cases and death. To prevent the morbidity and mortality rate caused by these diseases the only way to uptake vaccine. Despite evidence that vaccines are effective and safe, misconceptions persist. This can impact vaccine uptake and lead to lower levels of population vaccination coverage. Other information source like Media, Newspaper, News channel, health professionals have the significant impact to control the rate of vaccine refusal and hesitancy. Developing effective surveillance systems to track and monitor the spread of negative vaccine information may help to develop strategies to reduce the impact of exposure to negative information (Dyda, Shah et al. 2019).

Historically, methods to measure disease incidence, vaccination coverage and other health related outcomes have relied on information from health departments, registries, and large-scale surveys. Recent research uses information from the Internet for disease forecasting which helps to increase the transparency of health care system. Twitter produces a huge amount of data every day and it is first choice to the researcher because to work with public health data because of its public availability, social networking, as well as micro-blogging service. The author (Paul, Dredze et al. 2014) developed influenza forecasting health applications using data mining technique and the author (Mocanu, Baronchelli et al. 2013) developed indicator by measuring social belief and culture for specific community. The author (Eichstaedt, Schwartz et al. 2015) developed a tool for the sentiment and language analysis based on the Twitter data which measures hearth disease mortality rate in the geographical region (Eichstaedt, Schwartz et al. 2015). The rate of spreading the misinformation through social media, news channel, public opinion, and internet source (Salathé and Khandelwal 2011, Dunn, Leask et al. 2015) has become faster than ever and it directly impacts the public health (Oliver and Wood 2014). to take the right decision during vaccination program.

The misinformation which is related to health is extremely adverse and it influences people rapidly. Nowadays people are getting more influenced about health information through the online medium. They prefer to search rather than asking a health professional when they hear news about any news regarding health information like symptoms or vaccination. People justify health information whether it is fake, false true though internet source. Thereby, online medium has become vital factor to measure the public attitude and behaviour. To get rid of the curse of misinformation, a lot of research are going on to detect misinformation through online.

Vaccination beliefs and attitudes are an exemplar of the challenges of addressing misinformation in health. In a recent commentary, Larson (Larson 2018) suggests that the next major pandemic will be caused by viral misinformation. Persistent misinformed beliefs about vaccination have led to geographical clusters of refusal, which have been shown to increase the risk of outbreaks (Omer, Enger et al. 2008, Atwell, Van Otterloo et al. 2013). As countries improve access to healthcare services and reduce the impact of vaccine-preventable diseases, they see a corresponding increase in vaccine hesitancy and refusal (Larson, De Figueiredo et al. 2016).

In comparison to studies on misinformation in socio-political contexts, little empirical evidence is available for understanding vaccine-related information exposure through social media analyses. Previous work on human papillomavirus (HPV) vaccination has shown that differences in exposure to specific topics can explain how vaccination coverage varies from state to state in the United States (Dunn, Surian et al. 2017). Disclosure of misinformation regarding HPV vaccine tend to spread negative opinion (Dunn, Leask et al. 2015). For vaccination in general, there is a need to better understand how often people are exposed to low credibility information via social media, and whether there are populations of users who are more at risk of developing attitudes that lead to vaccine hesitancy and refusal because of the information to which they are exposed.

There are challenges associated with designing and measuring the effectiveness of interventions designed to mitigate the spread of misinformation and de-bias individuals because of the interplay between social and cognitive factors (Lewandowsky, Ecker et al. 2012), and this is certainly also the case for vaccination (Brewer, Chapman et al. 2017). Studies examining the effects of messaging interventions have produced mixed results (Nyhan, Reifler et al. 2014). Reviews of the trials that tested the efficacy of messaging interventions on attitudes and behaviours found that most trials failed to demonstrate efficacy (Sadaf, Richards et al. 2013, Fu, Bonhomme et al. 2014). Other studies in politics and related areas have also occasionally observed backlash, where participants presented with evidence that does not support their prior beliefs become further entrenched in their views and intentions (Flynn, Nyhan et al. 2017).

Most studies in the area evaluate the effects of interventions on individuals in isolation without considering the role of social network structure on the development and persistence of misinformation and attitudes that might lead to harmful behaviours. More research should be focused on the detection of misinformation in the social media as well as analysis the impact due to spread of misinformation (Dunn, Mandl et al. 2018).

## 2 Methods

**Study Data**

We worked on our previously collected twitter data which was related to vaccine. The duration of data collection was 17 January 2017 and 14 March 2018. For searching on the Twitter API (Application Programming Interface), we defined some search keywords earlier as (``vax*'',

``vaccin*'', ``antivax*'', ``immunis*'',). We extracted data from the tweets and stored the metadata and text. We only considered the English Language. We also collected the followers-following network of the users who posted these tweets. Finally, after removing the other language, we were able to download 6,591,566 tweets and retweets. In total, 1,860,662 users were involved for this tweets and retweets.

**Communities Detection**

We used a graph structure to represent the Twitter follower network where a node is a user and edges are formed from users to other users who they follow. For each user, we equally distributed the weight to all outgoing edges, so the total weight is 1. We then removed the following direction from the Twitter follower network and merged the edges and summed the edge weights between two users if they are following each other.

We ran community detection algorithm to find groups of users that each group has denser connections internally than the rest of the network. The Louvain algorithm was used as the algorithm can scale to large networks (Blondel, Guillaume et al. 2008). The algorithm is initialised with each node belongs to its own community and iteratively aggregated with the neighbouring community until no further modularity increasing reassignments of communities are produced.

As the Louvain algorithm relies upon the maximisation of modularity, it may result communities with many sizes, either too large or too small to be meaningful in the characterisation of communities. We heuristically determined that communities with more than 10,000 users were too large and needed to be processed further. To facilitate this, we sequentially ran the Louvain algorithm. Communities with more than 10,000 users resulted from the Louvain algorithm were separated from the network and the algorithm was run again to further divide the communities. From each run of the Louvain algorithm, we iteratively merged small communities until there were 10,000 users or less in each community. Two neighbouring small communities were merged if the total of edge weights between them was higher than the rest neighbouring communities. The process of dividing large communities and merging small communities was repeated until there were not any more changes in the community assignments. We used the implementation of the Louvain algorithm from python igraph[1].

**Characterising Communities**

In this study, we characterise online users' communities who are posting vaccine-related tweets using various measures. The first measure is to use the sharing of published articles. The aim of this measure is to see if communities can be distinguished from each other based on users sharing behaviour of published articles in different communities. So, for this, we separated the tweets in our study data into four categories based on the types of links exist in the tweets. Category 1 contains[2]. Category 2 contains tweets that include links to web pages (including news, blogs, Reddit, and Wikipedia), where the web pages have direct links to PubMed articles. These web pages were identified using [3] that captures the attention around scientific literature, which we queried using PubMed IDs of vaccine-related journal articles. Category 3 contains tweets that include links to Facebook, YouTube, Instagram and other social media posts. Category 4 contains tweets that do not include links to web pages.

The second measure is to use the sharing of low credibility information. The aim of this measure is to see if communities can be distinguished from each other based on how often users share low credibility information in different communities. For this, we need the credibility scores of the web pages, which makes sense for web pages in category 2 only. So, we selected the web pages in category 2 and downloaded the contents of those web pages. We considered the web pages that had at least than 300 words and we ignored the contents, links which were not in English. Also, we did not consider the web pages which were unavailable. Finally, we were able to extract 144,878 web pages which we used as category 2. Three investigators who were expertise in public health and communication domain, developed a seven-point checklist-based appraisal tool. The investigator estimated the credibility of these web pages through seven criteria. i) The source of information, which is presented should have scientific research background and objective. ii) The research should provide adequate detail information with evidence iii) The research should describe its drawback and uncertainties. 4) The research should not exaggerate, overstate information, and avoid the false evidence. 5) The research should be focused on context 6) The research should describe clear, understandable non-technical language 7) The research should acknowledge its sponsor if it is funded.

---

[1] http://igraph.org/python
[2] https://www.ncbi.nlm.nih.gov/pubmed
[3] https://www.altmetric.com

Finally, the investigators selected 474 vaccine related web pages. They applied the credibility appraisal tool based on the seven criteria. If any criteria were found to be matched, then they marked it 1 otherwise they marked it as 0. For example, if any research matched with 4 criteria and 3 did not match. and then it is calculated as (1+1+1+1+0+0+0=4). The research is measured as 4/7. To measure the rest of the web pages which are not related to research, we classified them using document classifier algorithm. We applied support vector machine and random forest to develop a new model which could predict the untrained model (unseen webpages) and to classify them. We trained 14 classifiers, one per criterion per machine learning algorithm (SVM and RF) using 474 manually expert-labelled documents. We considered web page as a document to represent them feature vectors where frequency represented the features vector. We assigned features vector weight according to the inverse document frequency (TF-IDF) for each unique word in the content. We measured the performance of the classifiers based on 10-fold cross-validation tests. We observed that if the web page contained the vaccine related information the credibility criteria score was more than 90% accuracy. The number of criteria satisfied by each web page is used to define a credibility score. We divided into three categories (low, medium, high) based upon the criteria satisfied for each webpage. We defined them low (0-2), medium (3-4), and high (5-7).

Table 1: The details of the measures used to characterise communities.

| Measures | Descriptions |
| --- | --- |
| Videos_pct | The percentage of the links with multimedia contents |
| Videos_avg_likes | Average likes for the links with multimedia contents |
| Low_cred_pct | The percentage of the links with low credibility contents |
| Low_cred_avg_likes | Average likes for the links with low credibility contents |
| High_cred_pct | The percentage of the links with high credibility contents |
| High_cred_avg_like | Average likes for the links with high credibility contents |
| Pub_articles_pct | The percentage of the links pointing to published articles |
| Pub_articles_avg_likes | Average likes for the links pointing to published articles |
| No_URLs_pct | The percentage of the tweets with no URLs |
| No_URLs_avg_likes | Average likes for the tweets with no URLs |
| Internal_density | The internal connection density of the users in the community |
| Comm_avg_likes | Average likes per community |
| Median_followers | Median number of followers per community |
| Users_avg_tweets | Avergae tweets per user in the community |

We have also used other measures (Table 1) to characterize communities, such as the average number of likes per tweet for all tweet types, the proportion of tweets with no URLs and the internal density of the communities etc. These measures are useful to characterize communities from various angles. For instance, the average number of likes across all types of tweets shows the popularity of the users in the community. The proportion of tweets with no URLs shows who is talking a lot. The internal density shows how well users are connected within the community. We have computed most of these measures from metadata of the tweets or users' social network.

Finally, we took these measures, found the rank percentile of each measure, and computed the deviation of the rank percentile of each measure to visualize communities.

# 3 Results

**Community structure**

To identify the structure of the community we applied the Louvain community detection algorithm along with the DMM values. The experimental result between community structure and the topics of tweets produced very well associated values which were generated from this algorithm. We applied the Louvain algorithm over 6,591,566 Twitter posts (tweets). 1,860,662 users tweeted these tweets. Based to the aligned values we formed 192 distinct communities.

We constructed community by joining small communities in an iterative process until the number of users were 10,000 users or less in each community. Two neighbouring small communities were merged if the total of edge weights between them was higher than the rest neighbouring communities. The group were formed based on the credibility scores of the web pages which were posted on twitter. The main difference between our work and others is that we measure creditability of web pages whereas other measure the credibility of tweets. The lower range of credibility score is 0-2, medium range of credibility score is 3-4 and higher range of credibility score is 5-7. The groups were formed semantically on the related topics hierarchical order based on the Louvain method and DMM.

**Topic Grouping**

To illustrates how the topics clusters within communities, we selected four topics and we visualized the measurement in a network constructive way where each community relates to other if there is relation among themselves. The size of the big circle indicates that the community has a vast number of users for the specific measurement. We coloured the circle according to gradient. The range of color is from red to green where red color indicates negative credibility and green color indicates the positive credibility.

Figure 1(a) shows category 1 which contains direct links to published articles with Digital Object Identifiers (DOIs) on PubMed. There are some large communities in category 1. But their credibility is extremely low. Most of them are red color. They do not share webpage in the posts among the community. Fig 1(b) shows low credibility among communities. The green color shows the percentage of sharing link. We can observe from Figure 1(c) most of the communities are green and rest are close to green and white. It depicts that in general communities prefer tweeting without URL sharing. Figure 1(d) represents the communities who share multimedia content link.

In figure 2 represents only 4 communities characteristics among 192 communities. The percentage of the sharing measurement is labelling from negative to positive range. Figure 2(a) represents the characterises of community 1. Figure 2(b) represents the characterises of community 18. It can be observed from figure 2(c) that community 95 are mostly active in every measurement. Figure 2(d) illustrates Community 101 are mostly inactive

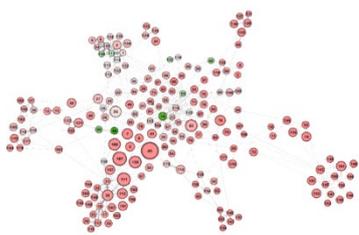
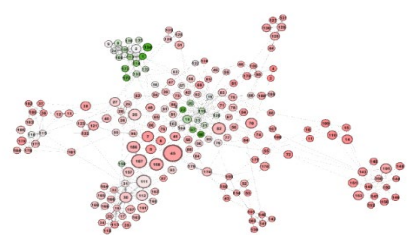

(a) Cat1 pct	(b) Low pct

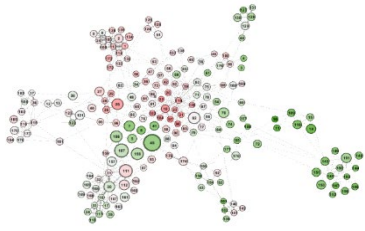

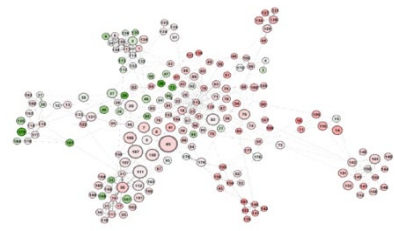

(c) No URLs pct            (d) Videos pct

Figure 1. The characterisations of communities using various measures.

Analysing the characterises of all communities we concluded in a discussion that users want to spread their own belief and experience inside their community regarding to vaccine. Majority tweets posted in a community were related to evidence and advocacy.

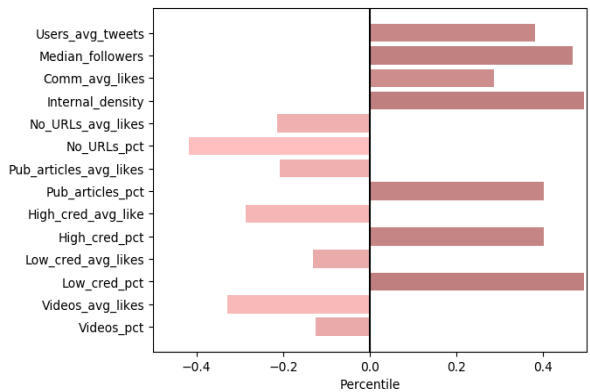

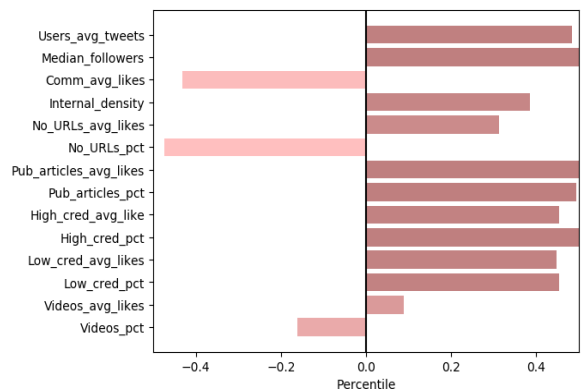

(a) Community1            (b) Community18

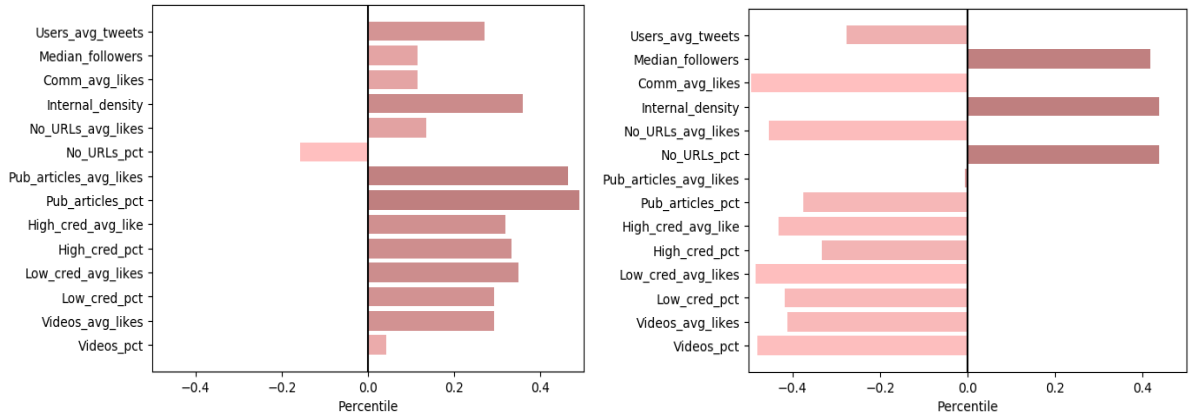

(c) Community 95      (d) Community101

Figure 2. The characterisations of communities using various measures

## 4 Discussion

**Principle Results**

In this study, we analysed the community structure as well as the network architecture among the community how they are connected to each other. We searched for the tweets regarding to vaccine information and who are posting inside community. We distinguished each community from one to another based on some measurement criteria. Then we followed up from the big community to small community opinion, posting, like, reaction. We also explicitly measured credibility of the link they shared among the community. We analysed the tweets of each community regarding the vaccination where users share opinion as well as their experience. This will help to identify the people perception regarding vaccination.

**Comparisons with Prior Work**

Several studies have been developed by the researcher to characterise the twitter users network structure. They classify the structure based on the sequences of tweets (Dunn, A.G et al. 2015) (Omer, S.B et al.2008). They formed cluster on the similar characteristics from the twitter data and make analysis over the connected network. In general, the prediction was made over the content of tweets (Larson, H.J et al. 2016) (Paul, M.G. Et al. 2014) Conversely, other used survey data to make prediction over the public atttude to health [Fu, L.Y et al. 2014] [Larson, H.J et al. 2016][Nayhan, B.J et al. 2014].

The differences with other studies and our work are that we characterise the community structure based on the link the users shared and other characterise based on the content. We identified the positive and negative community based on measure value.

**Limitations**

We used only twitter data to predict public attitude, behaviour over vaccination. To measure attitude of vaccine uptake by people which is critical decision should not be relayed only this evaluation. Also, we focused on the link shared by the users. Thereby, to evaluate public perception it needs both the content and the link.

**Future work**

We will be more focused on developing the characterises of community. We will try to combine with new spatial and demographic estimation methods to measure both the credibility of content and the shareable link together. It will determine the specific concern whether there are increased percentage of vaccine hesitancy, vaccine uptake or refusal.

# 5 Conclusion

In this study, we have proposed a new idea to evaluate the characteristics of community of the twitter user and their interaction through network architecture. We applied the Louvain community detection algorithm along with the DMM values to identify the community. We proposed a unique approach to measure community structure. We measure the credibility of webpages shared by the users using two machine learning algorithms namely support vector machines (SVM) and random forests (RF). The predicted result will help the public health organizations to take proper step to increase the public awareness for vaccination. It might be a potential step towards the development of automated public health surveillance systems in future.